\documentclass[preprint, aps, draft, pre]{revtex4}

\usepackage[dvips]{graphicx}

\begin{document}
\title{Organization of Block Copolymers
using NanoImprint Lithography: Comparison of Theory and Experiments}
\author{Xingkun Man$^{1}$, David Andelman$^{1,*}$, Henri Orland$^{2}$, Pascal Th\'ebault$^{3}$,
Pang-Hung Liu$^{3}$,
Patrick Guenoun$^{3}$, Jean Daillant$^{3}$, Stefan Landis$^{4}$}
\affiliation{
$^{1}$Raymond and Beverly Sackler School of Physics and Astronomy,
Tel Aviv University, Ramat Aviv 69978, Tel Aviv, Israel\\
$^{2}$Institut de Physique Th\'eorique, CE-Saclay, F-91191 Gif-sur-Yvette
Cedex, France\\
$^{3}$IRAMIS, LIONS, UMR SIS2M 3299 CEA-CNRS, CEA-Saclay, F-91191 Gif-sur-Yvette Cedex, France\\
$^{4}$CEA, LETI, Minatec, 17 rue des martyrs, F-38054, Grenoble Cedex 9, France}

\date{Jan 2, 2011}

\begin{abstract}
We present  NanoImprint lithography experiments and modeling of thin films of block
copolymers (BCP).
The NanoImprint technique is found to be an efficient
tool not only to align lamellar phases perpendicularly to the substrate,  but also to get rid of in-plane defects over
distances much larger than the natural lamellar periodicity.
The modeling relies on self-consistent
field calculations done in two- and three-dimensions, and is found to be in good agreement with the
experiments. It also offers some insight on the NanoImprint lithography setup and  on the conditions  required to
perfectly ordered BCP lamellae.
\end{abstract}

\maketitle

\baselineskip=18pt
\section{Introduction} \label {sec.1}

One of the main challenges of contemporary design of microchips is to find
affordable techniques of patterning silicon wafers at the nanoscopic level \cite{Bang09}.  To address this question,
self-assembling block copolymers (BCP) have been suggested as potent candidates  to provide patterns
for nanolithography \cite{Black07,Kim09}.
The aim is to produce thin and structured BCP films with patterns such as lamellae, cylinders and spheres~\cite{Leibler80} that can then be transferred to a substrate.

When thin films of BCP are cast on a surface,
they self-assemble into one of several possible nanostructures having
a specific orientation with respect
to the substrate.
In particular, by adjusting the surface interactions
and film thickness, it is possible to produce
lamellar and cylindrical phases in an orientation perpendicular~\cite{Mansky97,Liu09} to the substrate.
However, while lamellar and cylindrical phases can be perpendicularly aligned on a large
scale~\cite{Ham08,Han08}, the ever-remaining challenge for
micro-electronic applications is to find affordable and efficient techniques for in-plane organization,
with minimal amount of defects. This will allow producing devices that are hundreds of
micrometers in size, where precise spatial
accessibility is required.

Several attempts have been made to address this challenge. They include, among others,
chemical patterning of the substrate by e-beam lithography~\cite{Ruiz08,Stoykovich05} or by prepositioning of another
copolymer layer~\cite{Ruiz07}, and graphoepitaxy~\cite{Segalman01}, where an artificial surface
topography of grooves separated by walls is created on the substrate.
Due to such topographical constraints, ordered regions of BCP are
obtained over length scales of micrometers~\cite{Segalman01,Park07,Cheng02,Sundrani02}.


A more recent technique addressing the same issue is the nanoimprint lithography (NIL)\cite{Chou95,Hu05,Li02,Kim08}, and it has potential advantages in terms of cost and simplicity.
It uses surface micrometer-sized structural features of a reusable mold made by standard
lithographic techniques to guide the self-assembly of the BCP at the
nanometer scale.  In Ref.~\cite{Li02} a cylindrical phase oriented perpendicularly to the substrate
was imprinted by NIL where the NIL patterns are about twice or three times larger than the
BCP period. The resulting
BCP phase is aligned perpendicularly, but  the imprint procedure is found to induce unwanted defects.
In another attempt~\cite{Kim08}, alignment of lamellae was produced by NIL,
but the lamellae were not oriented perpendicular to the substrate.

Motivated by previous studies, we attempt in this paper to improve on the NIL procedure.
In particular, we focus on NIL setups that produce perpendicular aligned lamellae
with no defects, even when the NIL structural features are much larger than the BCP period.

\section{Materials and methods}

The symmetric di-block copolymer polystyrene-b-polymethylmethacrylate (PS$_{\rm 52K}$-b-PMMA$_{\rm 52K}$, PDI:1.09)
was purchased from Polymer Source Inc and exhibits a
lamellar phase of period $\ell_{0}$= 49\,nm in the
bulk~\cite{Stoykovich05}. PS and PMMA blocks share very similar
values of surface tensions, in the range of 29.7\,mN/m -- 29.9\,mN/m
and 29.9\,mN/m -- 31\,mN/m, respectively, at the temperatures used
for film annealing ($\sim$170$^\circ$C). The
glass transition temperature of PS and PMMA is 100$^\circ$C and
105$^\circ$C, respectively.

UV/ozone apparatus was used both for wafer cleaning and octadecyl-trichlorosilane (OTS)
self-assembly monolayer (SAM) oxidation. The wafers were cleaned by irradiation before the
silanization treatment. The OTS SAM oxidation was performed as described in Ref.~\cite{Liu09} in order to reach a precise surface
energy. To get a better contrast during AFM imaging, selective phase etching of PMMA can
be applied just before the imaging. To avoid ozone formation that would degrade both the PS
and PMMA blocks, the reaction chamber was flushed for 30\,min with nitrogen before and
during UV-irradiation. Samples were then irradiated for 30\,min and subsequently treated
with glacial acetic acid during 1h to ensure complete removal of degraded PMMA fragments.

\begin{figure}[h]
\begin{center}
\includegraphics[bb=0 0 470 201, scale=0.5,draft=false]{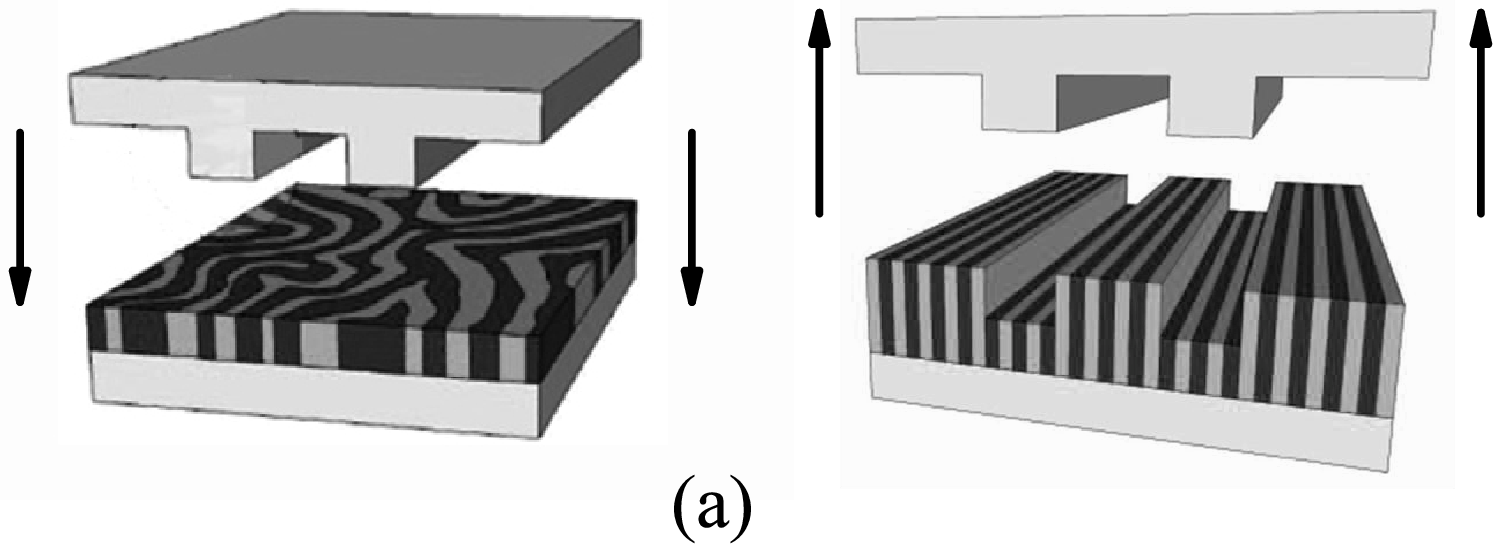}
{\includegraphics[bb=30 28 340 255, scale=0.5,draft=false]{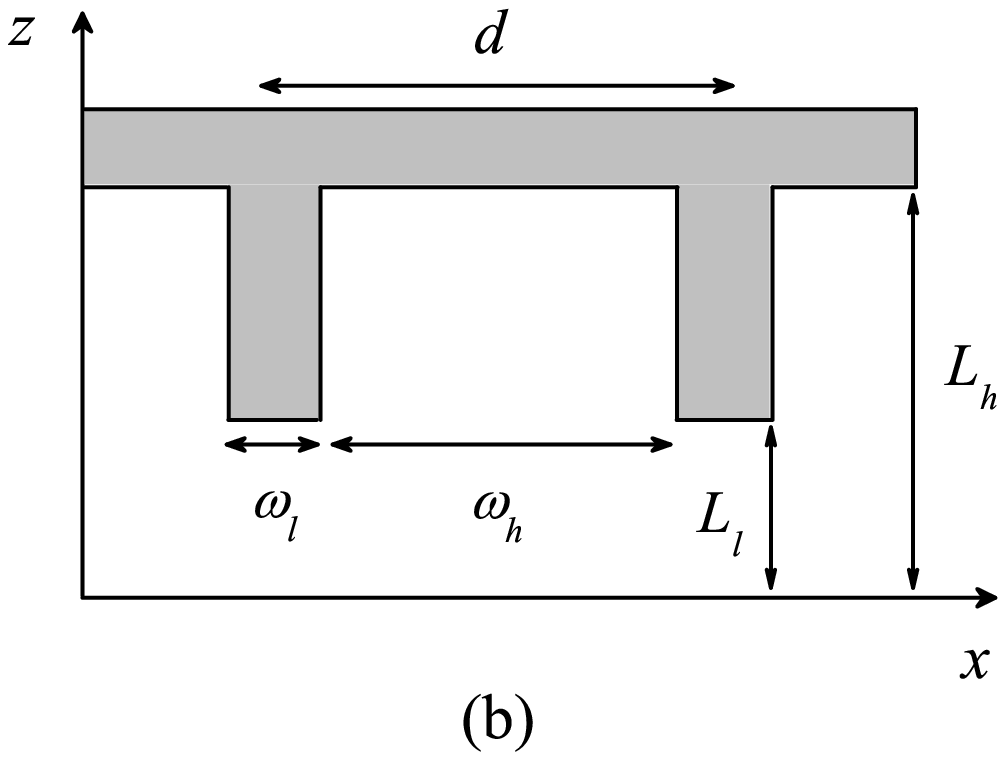}}
 \caption{\textsf{The NIL setup. (a) A grooved top mold is pressed upon a BCP film oriented perpendicularly to a
bottom substrate.  The lamellae follow the
direction of the mold grooves.  (b) A cut (side view) through the modeled NIL setup consisting of a
    periodically patterned mold (top surface).
    The periodicity in the $x$-direction is $d=\omega_l+\omega_h$,
    with $\omega_l$ and $\omega_h$ being the finger width and groove width,
    respectively. The BCP film fills the gap between the two surfaces and its
    thickness varies between $L_l$ and $L_h$. }} \label{fig1}
\end{center}
\end{figure}

Solutions of 1\,wt\%  of BCP in toluene
were prepared and spin coated at 1800\,rpm onto silanized silicon
wafers to produce BCP films with thickness of about 40\,nm (below
$\ell_0$). Subsequently, the samples were annealed in a vacuum oven
at 170$^\circ$C at a pressure of less than 3\,KPa for 1 day. The resulting lamellar phase was examined and
found to be perpendicular to the neutral substrate but with many in-plane defects~\cite{Liu09}.
Then, the sample was treated by thermal NIL, which consists of embossing the thin BCP film,
heated above its glass transition temperature, by a reusable mold made
of a series of grooves as is shown schematically
in Fig.~\ref{fig1}(a).

Imprint experiments have been carried out at the CEA/LETI clean room in
Grenoble, France, using a EVG{\textregistered}520HE press.
The mold is a 4" silicon wafer with topographical features made
of grooves of tens of nanometers in height with groove width ($\omega_h$) and
inter-groove separation ($\omega_l$)
in the range of hundreds of nanometers~[Fig.~\ref{fig1}(b)].
The mold surface is coated with a perfluorinated
polymer layer to avoid BCP adhesion. An overall pressure of about
$0.3$\,MPa was applied to the BCP film, first at a temperature of
120$^\circ$C during 7\,hours, then at  170$^\circ$C during
60\,hours.

Surface imaging after imposing NIL on the BCP
film was performed by SEM and AFM. SEM imaging was performed with a
Hitachi 9300 apparatus, operated at optimal voltages of about
500\,V. This voltage allows observation of BCP phase organization, without PMMA etching, due to
a smaller penetration depth of the electron beam and enhanced
secondary electron emission yield. AFM imaging has been performed
using a NanoScope V (Digital Instruments). The samples were analyzed
using the AFM tapping mode in air and with silicon cantilevers of
125\,$\mu$m in length (Ultrasharp, Micro Masch). Their resonance
frequency ranges between 265 and 400 KHz, whereas their force
constant lies in between 20 - 75 N/m and the tip radius of curvature
is less than 10\,nm. The scan rate was chosen in order to obtain the
best contrast in phase images and the least deviation between height
trace and retrace scans.

\section{The Self-Consistent Field theory}
We use self-consistent field (SCF) theory to investigate the
behavior of a melt of A-B di-block copolymer (BCP) film at
nano-patterned surfaces. The BCP film has $n$ polymer chains, each
having a length $N=N_A+N_B$ in terms of the Kuhn length $a$, which is
assumed, for simplicity, to be the same for the $A$ and $B$
monomers. Hence, the A-monomer molar fraction $f=N_A/N$ is equal to
its volume fraction. In addition, hereafter we concentrate on
symmetric di-BCP, $N_A=N_B$ having $f=0.5$. The symmetric BCP yields
thermodynamically stable lamellar phases of periodicity $\ell_0$, as
temperature is lowered below the order-disorder temperature
(ODT). We rescale all lengths, $r\to r/\ell_{0}$,
by the natural periodicity of the BCP, $\ell_0\simeq 4.05R_g$, where $R_g$
is the chain radius of gyration $R_g^2=Na^2/6$. Similarly, the
curvilinear coordinate along the chain contour, $s$, is rescaled by
$N$, yielding $s\to s/N$. With these
conventions, the free energy for such a BCP film
confined between the two surfaces is

\begin{eqnarray}\label{f1}
\frac{N a^2}{\Omega_1\ell^{3}_{0}}\frac{F }{k_{B}T} &=&\frac{F}{n
k_{B}T}=\frac{1}{\Omega_1}\int {\rm d}^{3} r
\left[\chi\phi_{A}(r)\phi_{B}(r)-\omega_{A}(r)\phi_{A}(r)-\omega_{B}(r)\phi_{B}(r)\right]
\nonumber\\
&-& \ln Q_{C}-\frac{1}{\Omega_1}\int {\rm d}^3r
\left[u_{A}(r)\phi_{A}(r)+u_{B}(r)\phi_{B}(r)\right]
\nonumber\\
&  + &\frac{1}{\Omega_1}\int {\rm d}^3 r\, \eta(r)[\phi_A(r)+\phi_B(r)-1]
 \end{eqnarray}

The system has a total rescaled volume $\Omega_1$. The closed-packing density (monomer per unit
area) is $\rho_{0}=a^{-3}$, and the Flory-Huggins parameter is
$\chi$. The dimensionless volume fractions of the two components are
defined as $\phi_{A}(r)$ and $\phi_B(r)$,
respectively, whereas $\omega_{j}(r)$, $j={\rm A, B}$, are the
auxiliary fields coupled with $\phi_{j}(r)$, and $Q_{C}$ is the
single-chain partition function in the presence of the $\omega_A$
and $\omega_B$ fields. The third term represents a surface-energy
preference, where $u_A$ and $u_B$ are the short-range interaction
parameters of the surface with the A and B monomers, respectively.
Formally, $u_A(r)$ and $u_B(r)$ are surface fields and have non-zero values only
on the surface(s). It is worth noting that due to the above  rescaling,
the variables in Eq.~(\ref{f1}) should also be rescaled by:
$\chi\to N
\chi$, $\omega_{j}(r)\to N\omega_{j}(r)$ and $u_{j}(r)\to N
u_{j}(r)$.

Finally, the last term in
Eq.~(\ref{f1}) includes a Lagrange multiplier $\eta(r)$
introduced to ensure the incompressibility condition of the BCP melt:
\begin{equation}
\phi_{A}(r)+\phi_{B}(r)= 1 \quad {\rm for~~ all} \quad r\in \Omega_1
\label{f2a}
\end{equation}
By inserting this condition, Eq.~(\ref{f2a}), in the surface free
energy term of Eq.~(\ref{f1}), the integrand becomes $u_A\phi_A+u_B\phi_B=(u_A-u_B)\phi_A
+u_B$. Hence, $u(r)\equiv u_A(r)-u_B(r)$ is the only needed
surface preference field to be employed hereafter.

Using the saddle-point approximation, we can obtain a set of
self-consistent equations
\begin {eqnarray}
\omega_{A}(r)&=&\chi\phi_{B}(r)-u_{A}(r)+\eta(r)\label{f3}\\
\omega_{B}(r)&=&\chi\phi_{A}(r)-u_{B}(r)+\eta(r)\label{f4}\\
\phi_{A}(r)&=&\frac{1}{Q_{C}}\int^{f}_{0} {\rm d} s\,q_{A}(r,s)q^{\dag}_{A}(r,f-s)\label{f5}\\
\phi_{B}(r)&=&\frac{1}{Q_{C}}\int^{1-f}_{0}
{\rm d} s\,q_{B}(r,s)q^{\dag}_{B}(r,1-f-s)\label{f6}
\end {eqnarray}
where the incompressibility condition, Eq.~(\ref{f2a}), is obeyed,
$f=N_{A}/N$ and the single-chain free energy $Q_c$ is:
\begin{equation}\label{f7}
Q_{C}=\frac{1}{\Omega_1}\int {\rm d}^3r \,q^{\dag}_{A}(r,f)
\end{equation}
The two types of propagators $q_{j}\left(r,s\right)$ and
$q^{\dag}_{j}\left(r,s\right)$ (with $j={\rm A, B}$) are solutions
of the modified diffusion equation
\begin {equation} \label{f8}
\frac{\partial q_{j}(r,s)}{\partial s}
=\left(\frac{R_g}{\ell_0}\right)^2\nabla^{2}q_{j}\left(r,s\right)-\omega_{j}(r)q_{j}(r,s)
\end {equation}
with the initial condition
$q_{A}(r,s{=}0)\,{=}\,q_B(r,s{=}0)\,{=}\,1$,
$q^{\dag}_{A}(r,s{=}0)\,{=}\,q_{B}(r,1{-}f)$ and
$q^{\dag}_{B}(r,s{=}0)=q_{A}(r,f)$. This diffusion equation is
solved using reflecting boundary conditions at the two confining
surfaces  ($z{=}0$ and $z{=}L$): ${\rm d} q/{\rm d} r|_{z=0}\,{=}\,0$ and ${\rm d} q/{\rm d}
r|_{z=L}\,{=}\,0$, while  periodic boundary conditions are used in
the perpendicular direction.

The top surface with parallel and elongated grooves is modeled as an impenetrable boundary
for the polymer chains. In practice, this is done by assigning a large value to the
local surface field, $u_A(r)=u_B(r)=-20$, for all points inside the grooves. In addition, on the whole top surface we include a weak attractive field ($u=0.02$), which takes into account the preferred surface interaction with the A block.

We use the same numerical method as that in our previous work \cite{Manxk10} to
solve these self-consistent equations. Fields and densities were calculated on a grid with a spatial resolution of $\Delta \approx \ell_0/10$ or $\Delta \approx \ell_0/20$ for the three- and two-dimensional grids, respectively. First, we guess an initial
set of values for the $\{\omega_{j}(r)\}$ auxiliary fields. Then,
through the diffusion equations we calculate the propagators, $q_j$
and $\{q^\dagger_j\}$, and from Eq.~(\ref{f5}) and Eq.~(\ref{f6}) we
calculate the monomer volume fractions $\{\phi_j\}$. Next, a new set of
values for $\{\omega_{j}(r)\}$ is obtained through Eq.~(\ref{f3}) and
Eq.~(\ref{f4}) and this procedure can be iterated until convergence is
obtained by some conventional criterion. More details are given in Ref.~\cite{Manxk10}.

\section{Results}
\begin{figure}[h]
\begin{center}
\includegraphics[bb=50 70 305 239,scale=0.85,draft=false]{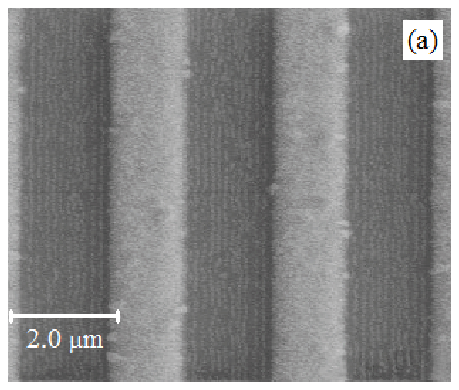}
\includegraphics[bb=0 0 339 335,scale=0.3,draft=false]{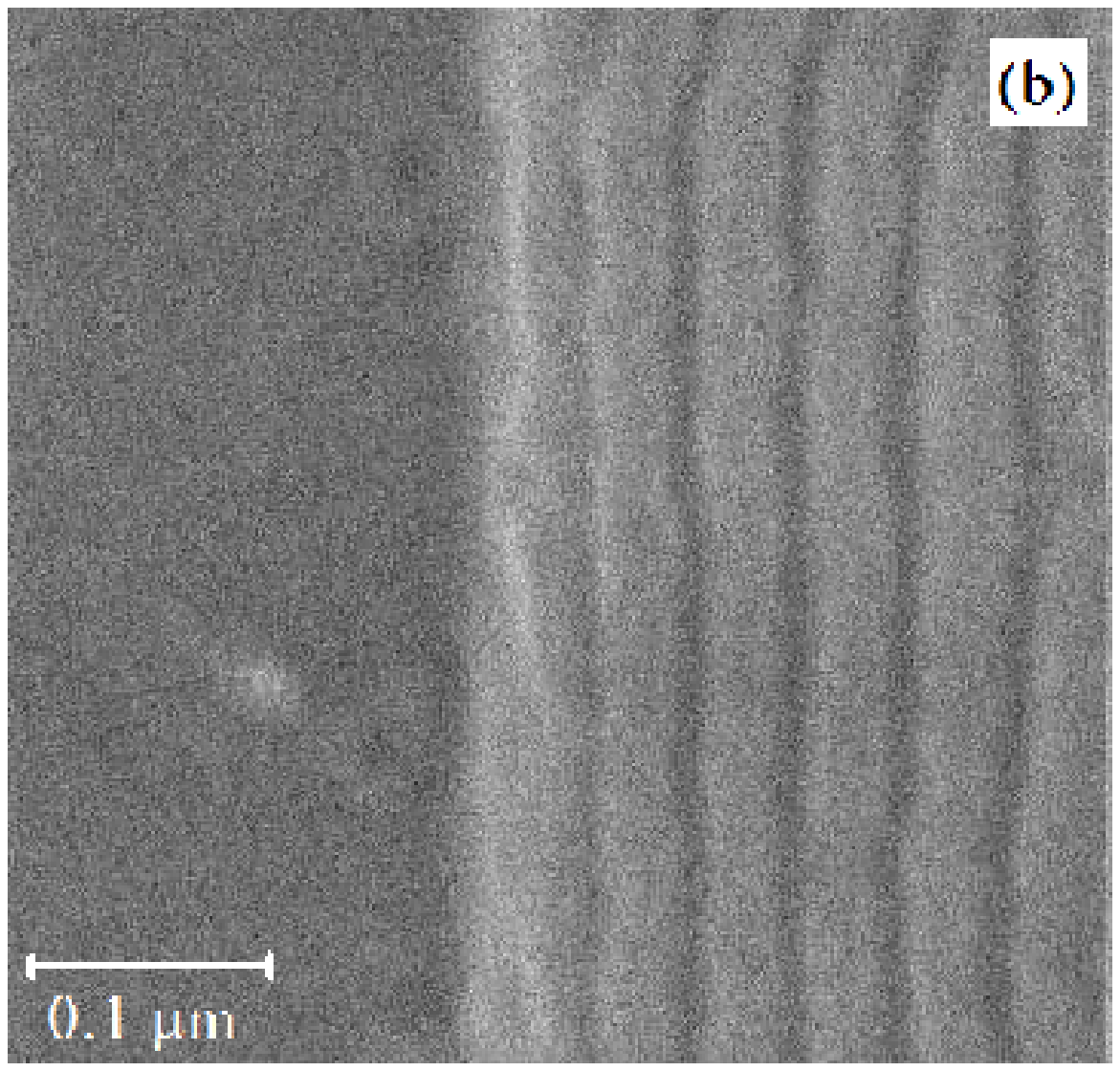}
\includegraphics[bb=50 62 305 260,scale=0.85,draft=false]{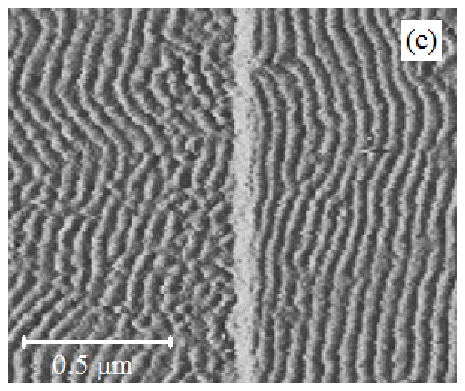}
\caption{\textsf{(a) Top view of a SEM image of the
BCP film after nanoimprinting with a NIL mold with $\omega_l=\omega_h=1.5$\,$\mu$m. The groove height
is $L_h-L_l=50$\,nm.
The perpendicularly lying BCP lamellae are further oriented
along the groove long axis. (b) An enlargement of (a) where several lamellae (left) are well ordered in-plane.
(c) An AFM top view of an enlarged section of the BCP film close to
thin-thick boundary (the middle vertical line).
To the right, the BCP film is thicker (thickness $L_h$) and is ordered by the groove vertical wall, whereas
to the left, the BCP is thinner (thickness $L_l$) and less ordered.}} \label{fig2}
\end{center}
\end{figure}

Figure~\ref{fig2} shows
a top view of SEM and AFM images of a PS$_{\rm 52K}$-b-PMMA$_{\rm 52K}$ thin film after nanoimprinting with a mold made of a series of parallel grooves [Fig.~\ref{fig1}]. From the images, it can be seen that the lamellar phase is indeed
oriented perpendicular to the bottom substrate. Moreover, in Fig.~\ref{fig2}(a) the SEM image indicates that the thicker sections of the BCP film exhibit lamellae that are nicely oriented
in a parallel fashion along the groove vertical walls (the $y$-direction), while the thinner sections of the film
are too thin to be visualized by the SEM technique. The same in-plane ordered lamellae are shown as an enlargement in Fig.~\ref{fig2}(b), whereas in Fig.~\ref{fig2}(c) the region close to the groove vertical wall is visualized by an AFM technique. The AFM image indicates a different in-plane ordering of the thick and thin sections of the BCP film. While the thick section (right) shows well ordered lamellae of periodicity very close to the bulk one, $\ell_0$, which are aligned by the wall, the lamellae of the thin section (left) of the film are less oriented and contain some defects. This observation can be attributed to the preferred interaction of the mold vertical walls with one of the two BCP
blocks. The lamellar film to the right of the wall is thicker in the $z$-direction
(height $L_h$ as in Fig.~\ref{fig1}(b)) and  is in contact with the vertical groove wall, while the film section on the left has
a smaller height ($L_l$), and the wall has no direct influence on it.

In order to further understand these NIL results, we complement the experiments with
self-consistent field (SCF) theory calculations performed on
symmetric BCP lamellar phases~\cite{Matsen97,Petera98,Pereira98,Geisinger99,Tsori01}. The NIL setup is modeled by
BCP lamellae having a natural periodicity $\ell_0$, and
confined in the $z$-direction
between a flat and neutral bottom surface ($u=0$) at $z{=}0$ and a topographical varying surface (the top mold), as is shown schematically in Fig.~\ref{fig1}(b).
The top surface
has the form of elongated grooves (in the $y$-direction)
of square cross section (in the $x$-direction). The down-pointing indentations (fingers) have a width of
$\omega_l$ separated by grooves with a cross-section width $\omega_h$.
The BCP film thickness measured with respect to the $z{=}0$ surface varies between
$L_h$ inside the grooves and $L_l$ in between the
grooves [Fig.~\ref{fig1}(b)]. The  mold preferential interaction toward one of the two blocks is modeled
by an overall non-zero surface field, $u >0$.

\begin{figure}[h]
\begin{center}
{\includegraphics[bb=0 -10 360 236, scale=0.5,draft=false]{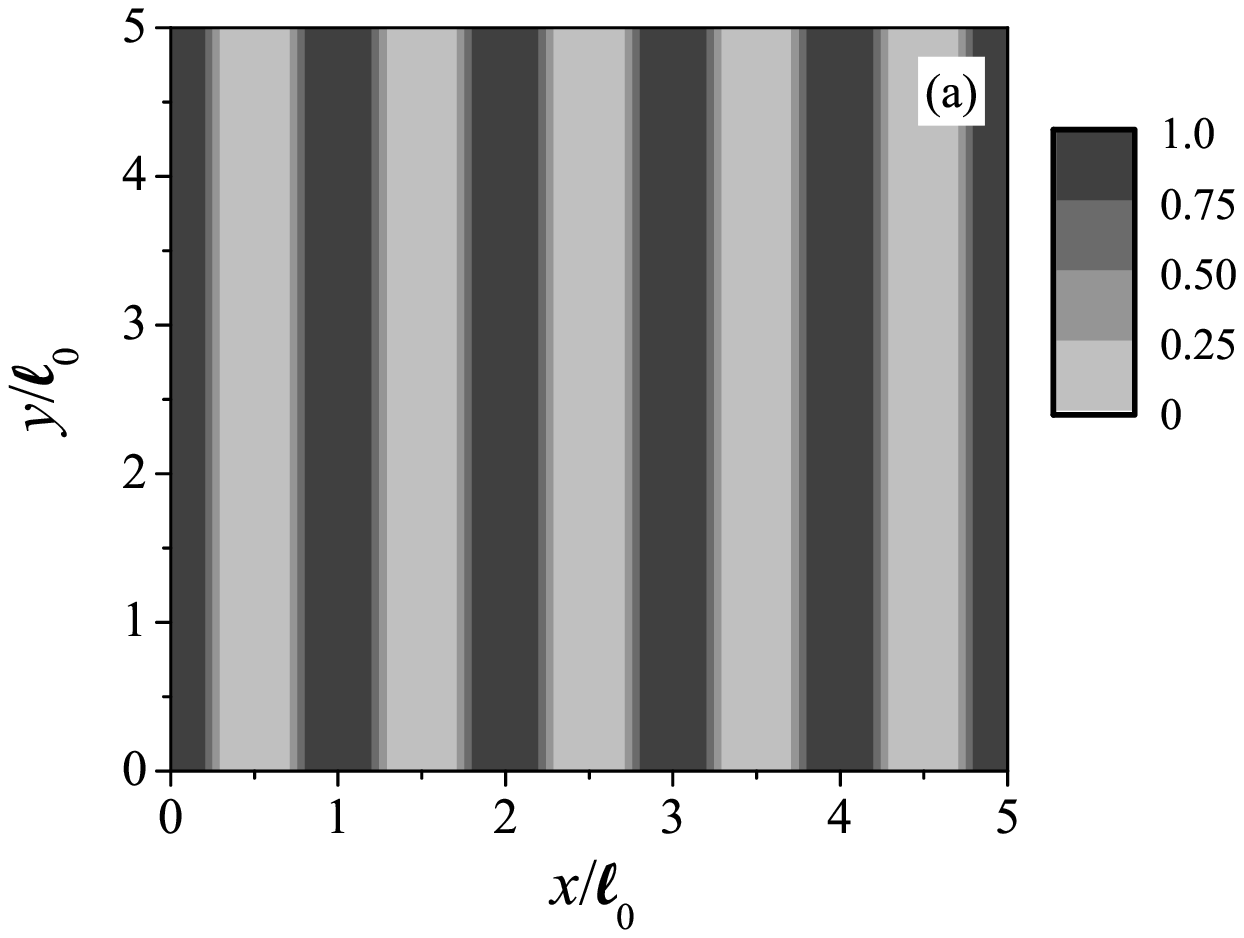}}
{\includegraphics[bb=0 -10 314 236, scale=0.5,draft=false]{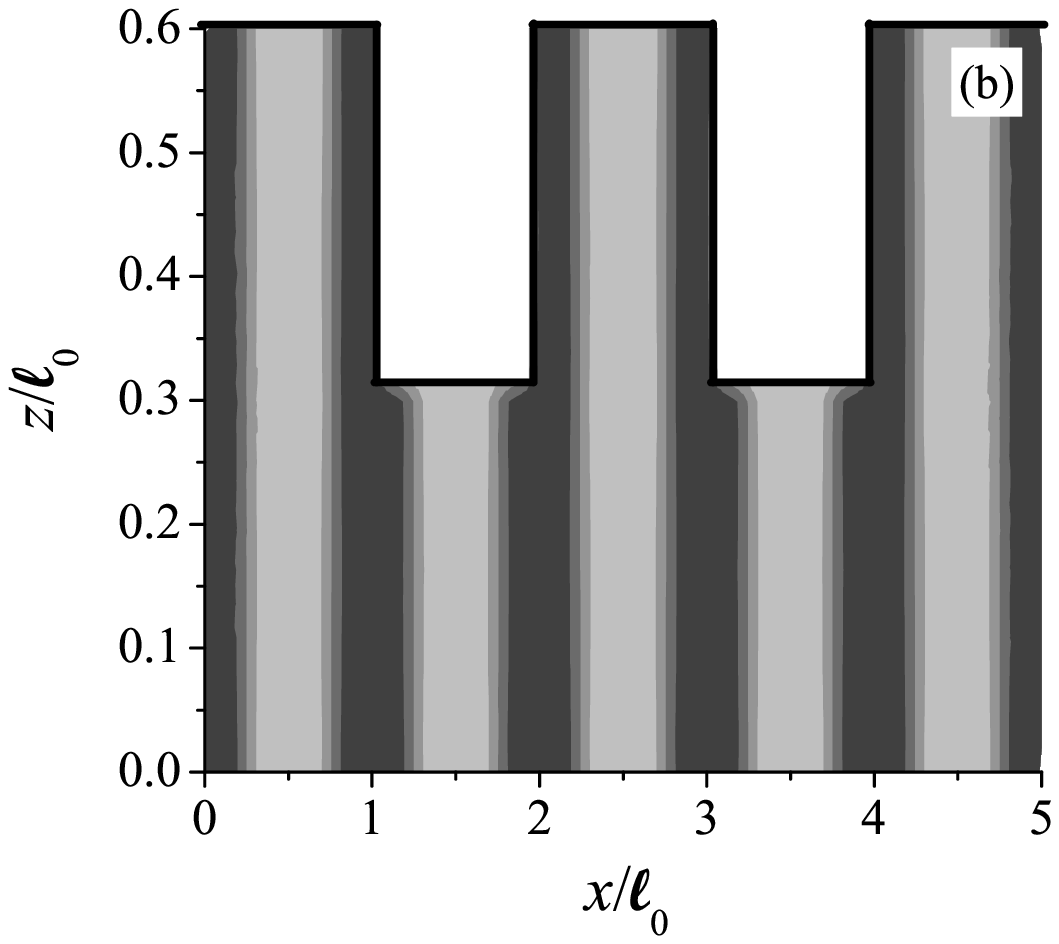}}
\includegraphics[bb=0 0 314 236, scale=0.5,draft=false]{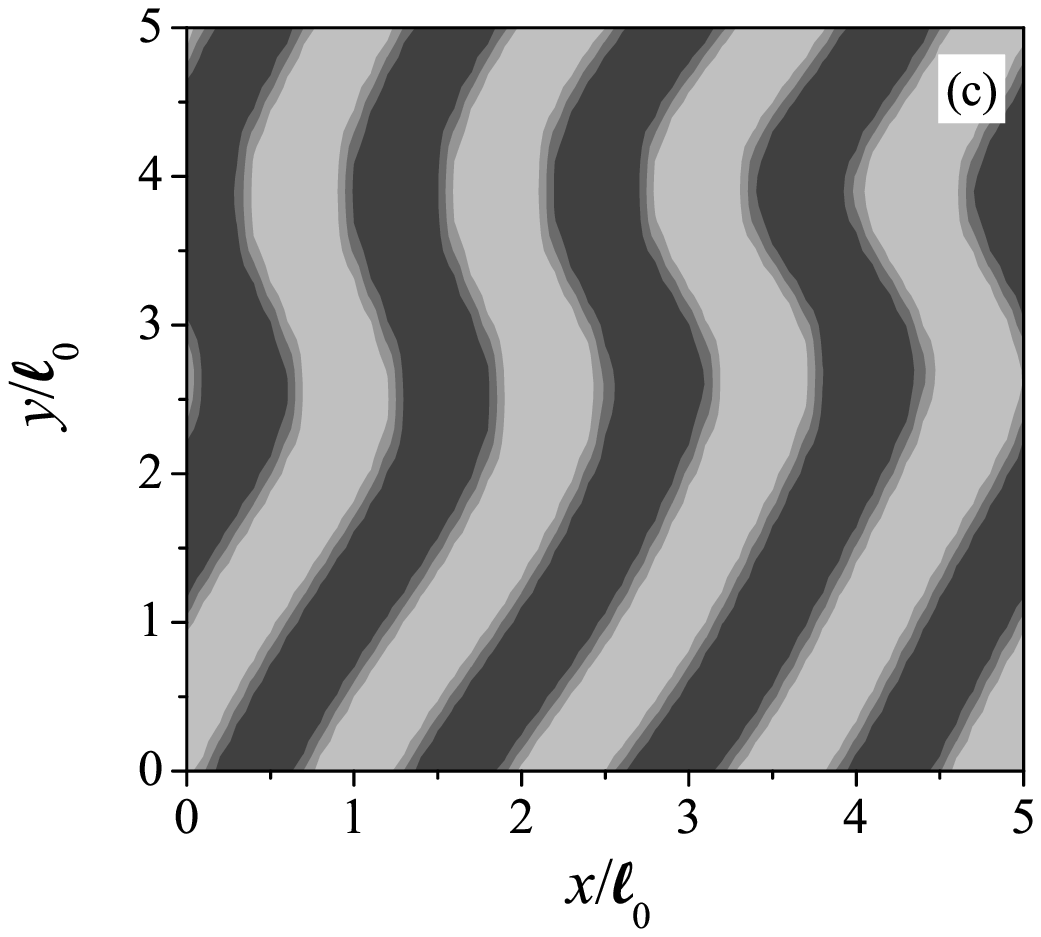}
\caption{\textsf{Calculated 3d lamellae in a BCP film of size:
$L_x\times L_y \times L_z=10\ell_0\times 10\ell_0 \times 0.6\ell_0$ and $N\chi=25.0$. The system parameters for the NIL are: $u=0.02$, $d=4\ell_0$,
    $\omega_l=\omega_h=2\ell_0$, $L_h=0.6\ell_0$ and
    $L_l=0.3\ell_0$. (a) Top-view cut at $z=0.3\ell_0$
and (b) side cut at $y=5\ell_0$ of the simulated NIL setup.  The perfect in-plane order seen in (a) is accompanied by
a perfect perpendicular order in (b). For comparison, a top-view cut at $z=0.3\ell_0$ is presented in (c) for
a similar system bound by two neutral and flat surfaces at $z=0$ and $z=0.6\ell_0$. It shows several in-plane defects embedded into
the perpendicularly oriented lamellar phase.}} \label{fig3}
\end{center}
\end{figure}

We performed several three dimensional (3d) SCF calculations to shed light on the film in-plane ordering. The 3d system size is $L_x\times L_y\times L_z=10\ell_0\times 10\ell_0\times 0.6\ell_0$, where $L_z$ is set to be less than one BCP periodicity in accord
with the experiments (Fig.~\ref{fig2}).
In our 3d SCF calculations, a gradual temperature quench is performed. The starting temperature is above the order-disorder temperature (ODT), $N\chi_c\simeq 10.5$; hence, inside the
BCP disordered phase. The temperature is then gradually decreased (or, equivalently, the value of $N\chi$ is gradually increased)
until the system reaches $N\chi=25$, which is well below the ODT.

\begin{figure}[h]
\begin{center}
{\includegraphics[bb=0 0 290 240, scale=0.65,draft=false]{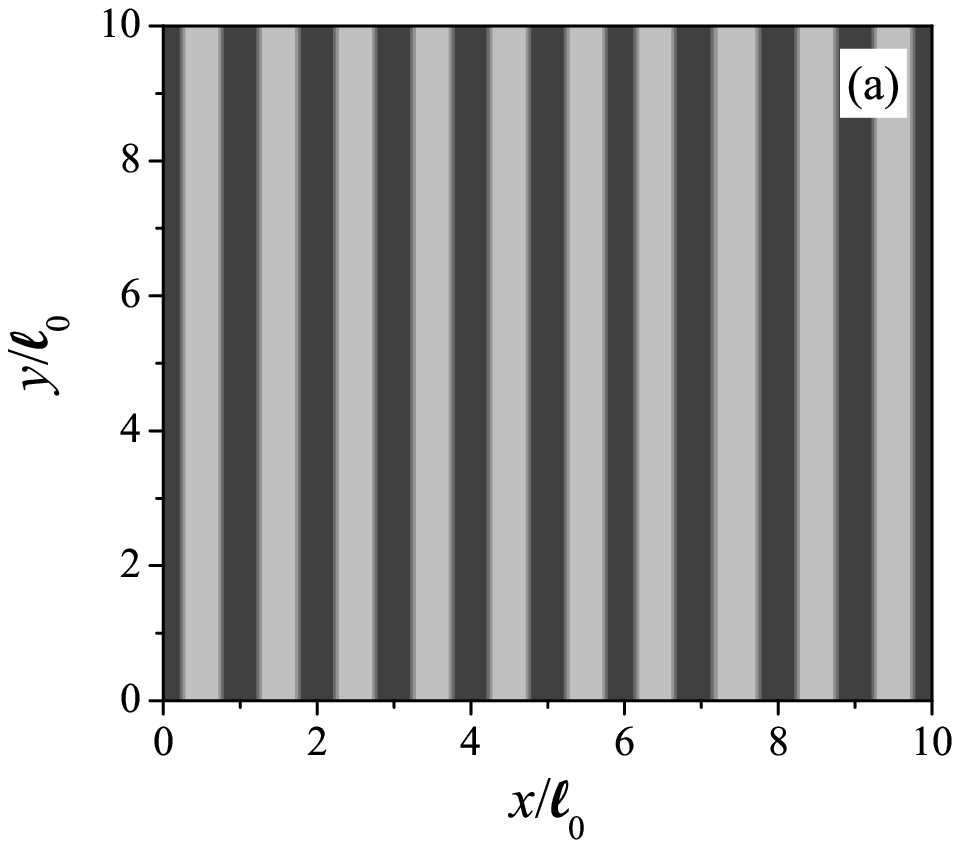}}
{\includegraphics[bb=0 0 390 245, scale=0.65,draft=false]{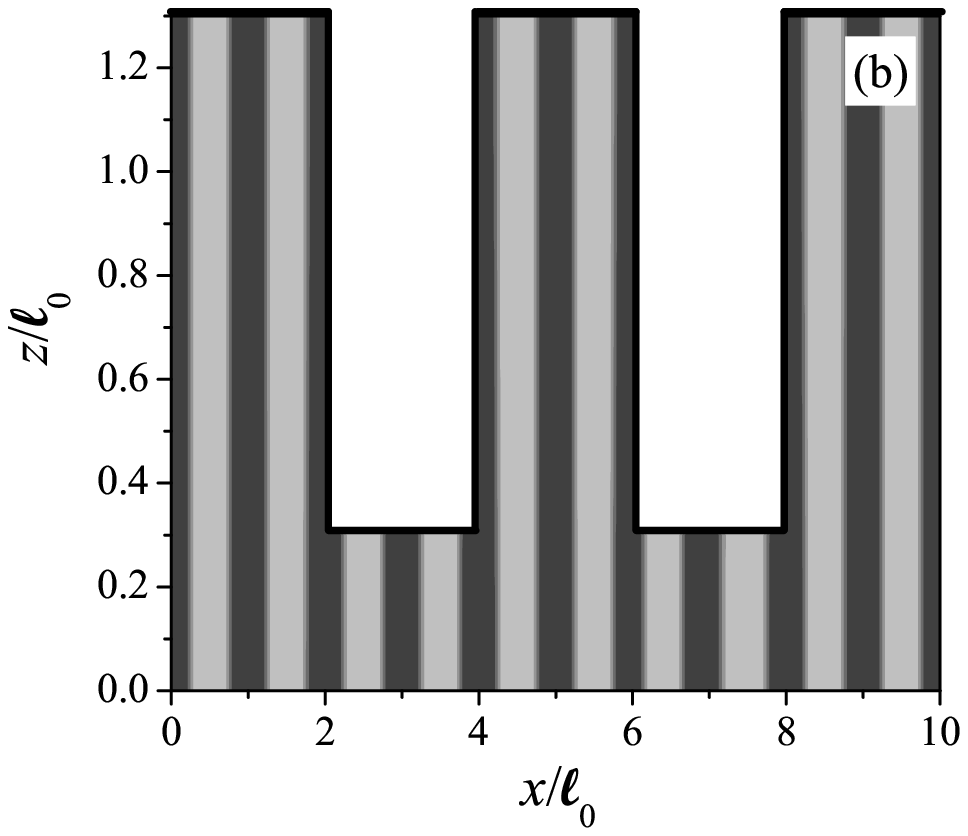}}
\caption{\textsf{Calculated 3d lamellae in a BCP film of size:
$L_x\times L_y \times L_z=10\ell_0\times 10\ell_0 \times 1.3\ell_0$. All other parameters are the same as in Fig.3(a). (a) Top-view cut at $z=0.3\ell_0$
and (b) side cut at $y=5\ell_0$ of the simulated NIL setup.  The perfect in-plane order seen in (a) is accompanied by
a perfect perpendicular order in (b).}} \label{fig5}
\end{center}
\end{figure}

\begin{figure}[htp]
  \begin{center}
    {\includegraphics[bb=0 0 308 270,scale=0.5,angle=0,draft=false]{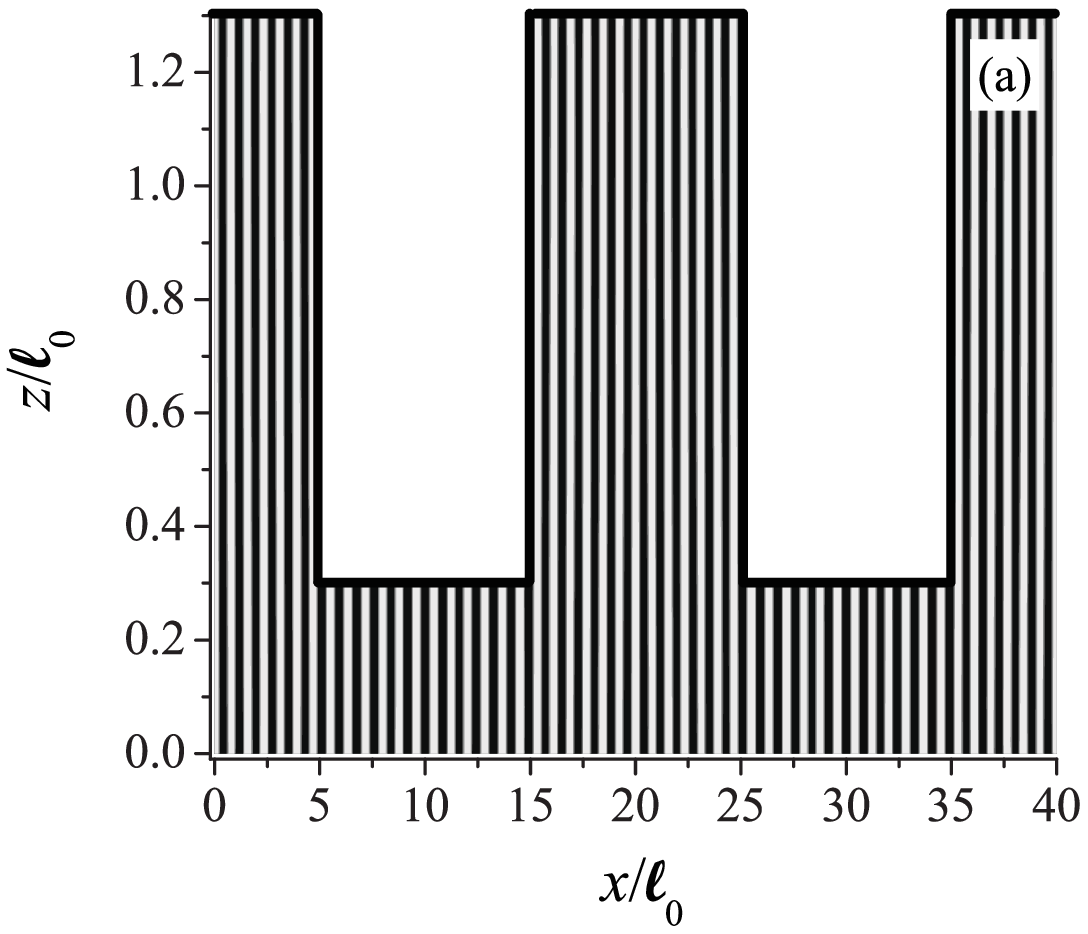}}
    {\includegraphics[bb=0 0 308 270,scale=0.5,angle=0,draft=false]{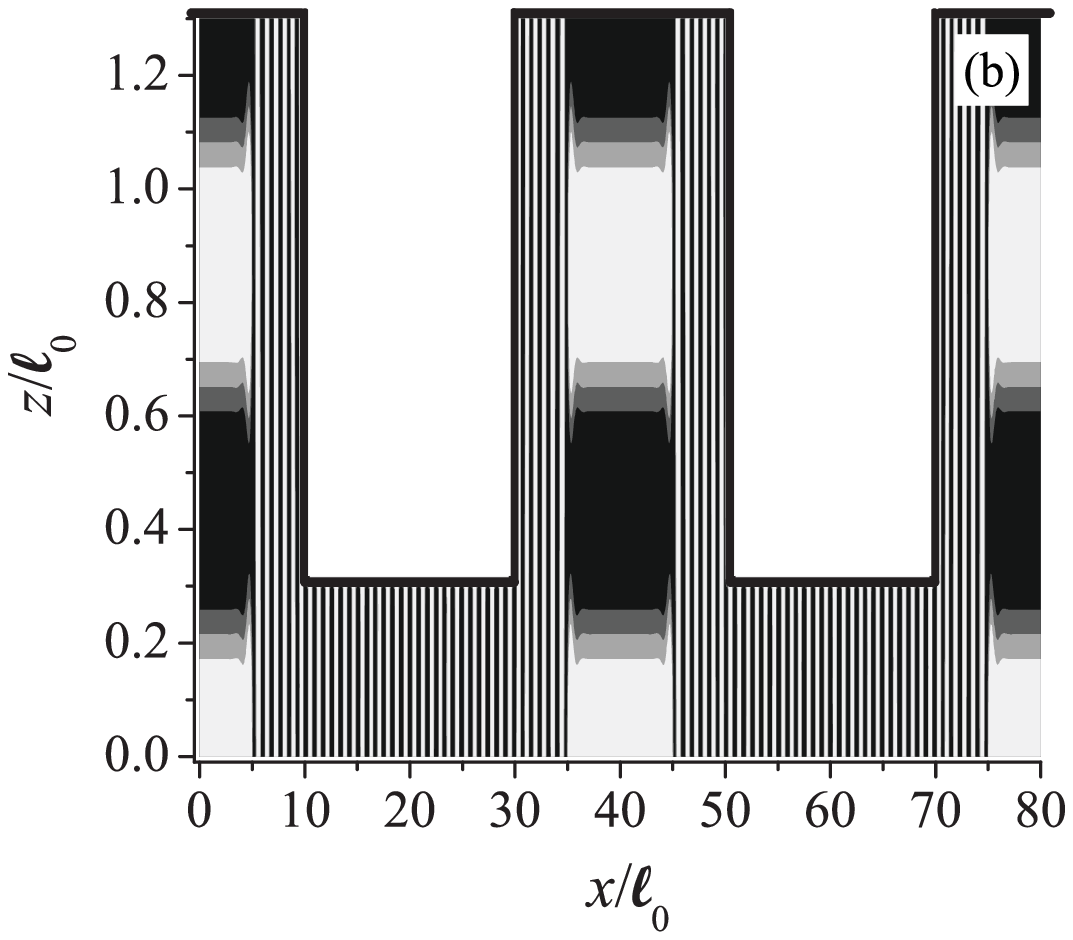}}
    {\includegraphics[bb=23 15 207 200,scale=0.73,draft=false]{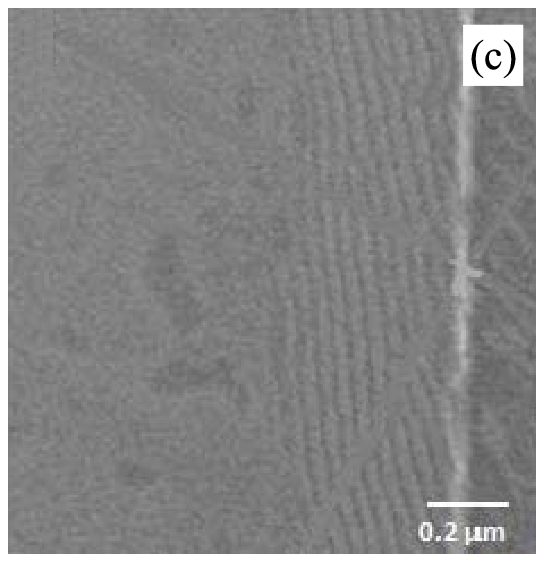}}
    \caption{\textsf{(a) Calculated  2d structures of BCP lamellae  in the $x{-}z$ plane for
     $N\chi=20.0$.  The system parameters are $u=0.02$, $d=20\ell_0$,
    $\omega_l=\omega_h=10\ell_0$, $L_h=1.3\ell_0$ and
    $L_l=0.3\ell_0$.
    (b) Same as (a) but with an increased groove periodicity
    $d=40\ell_0$ with $\omega_l=\omega_h=20\ell_0$.
    (c) Top view of a SEM image of a PS$_{\rm 52K}$-b-PMMA$_{\rm 52K}$
film after NIL.
For groove width of about $20\ell_0$ the perpendicular lamellar phase persists only near
the groove edge (white vertical line on the right). In the region far from the groove edge,
 the lamellae probably change their orientation to one that is parallel
 to the substrate. The bar scale is  0.2\,$\mu$m.
 \label{fig4}}}
  \end{center}
\end{figure}

With the simulated NIL mold,
a perfect perpendicular lamellar structure is found and is shown in Fig.~\ref{fig3}(a) as a top-view cut at $z=0.3\ell_0$, and in \ref{fig3}(b) as a side cut at $y=5\ell_0$. Note that in the calculation, Fig.~\ref{fig3}(a), the thinner and thicker  sections of the film are equally ordered, whereas in the experiments, Fig.~\ref{fig2}(c), the in-plane ordering is not as good for the thinner section of the film (left side of the figure) as compared with the thicker film section on the right side. This lack of in-plane ordering in the experiments might be due to lack of equilibration or shear effects at the walls, and
more detailed investigations are needed to further clarify this point.

As a further check, we compare the NIL setup with a BCP film confined between two neutral and flat surfaces. As can be seen in Fig.~\ref{fig3}(c), a perpendicular lamellae structure arises with no in-plane ordering when the same gradual temperature quench process is repeated for the neutral and flat surfaces. It is clear that there are many in-plane defects in this case, in contrast with the NIL mold that induces a strong in-plane ordering.

We show another 3d calculation in Fig. \ref{fig5}, for which all the parameters are the same as in Fig.~\ref{fig3} (a) except that we modify the $L_h$ value to be $L_h=1.3\ell_0$. We still obtain the perfect perpendicular lamellae for this thicker BCP film, which is consistent with the phase diagram of Ref~\cite{Manxk10}. We conclude that the value of $L_h$ is not strictly limited, but can have a range of values that will result in a perfect BCP perpendicular lamellar phase.

Another important question we addressed
is the stability of the perpendicular phase for wider grooves (larger $\omega_h$).
To further explore this issue we complemented
the 3d calculations by 2d ones where much larger lateral system
sizes (up to $L_x\times L_z=1.3\ell_0\times 80\ell_0$) can be simulated. Although the two-dimensional calculation cannot infer on the degree of in-plane ordering and defects, it is of value because it shows the effects of groove width on the stability of the perpendicular BCP lamellae.
In Fig.~\ref{fig4}(a), an ordered perpendicular phase is
found for the NIL setup with $d=20\ell_0$. This figure should be compared with the 3d
calculations as shown in Fig.~\ref{fig3}(a) for much narrower grooves of $d=2\ell_0$.

The NIL setup has some limitations as can be seen from our calculations
and through preliminary experimental results.
If we increase the groove periodicity (scaling up both $\omega_h$ and $\omega_l$),
the perfect perpendicular lamellae do
not persist. Instead, the film breaks up into a mixed morphology, combining
a perpendicular phase close to the groove wall with a parallel phase induced by the
horizontal section of the top surface. This can be seen in Fig.~\ref{fig4}(b), where all parameters are the same as in \ref{fig4}(a) beside
a larger groove periodicity, $d=40\ell_0$. These findings are
in accord with the experimental ones shown in Fig.~\ref{fig4}(c),
where SEM images demonstrate the loss of the perpendicular lamellae,
for large enough groove periodicity,
after distances of about $10\ell_0$ from the groove wall.

\section{Discussion \& Conclusions}

We address in this article the influence of nano-patterning of surfaces on the
orientation and alignment of lamellar phases
of BCP films, using a  NIL technique to produce superior perpendicular ordering.
The main goal of the NIL
is to be able to use surface features on length scales larger
than the BCP natural periodicity $\ell_0$ in order to reduce the cost of
expensive surface preparation treatments.

The effect of the NIL setup on BCP in-plane ordering is clearly demonstrated both in the experiments and in the modeling.
Without the NIL mold, it is possible to obtain perpendicular lamellae but with many in-plane defects
which cannot be annealed away. However, with the NIL mold, wetting of the vertical groove wall induces
perfect perpendicular ordering with minimal amount of defects, over large lateral distances.

In our model the perfect perpendicular order is induced
by a small surface field $u=0.02$, chosen to
agree with PS-PMMA experimental setup, where the relative difference in the surface tension
between the two blocks  and the surface is about 1\%.
The value of  $u$ is limited by two opposing trends. On the one hand, $u$ should
be large enough so that one of the two blocks would wet the groove vertical wall. On the other hand,
$u$ should be small enough in order not to interfere with the overall perpendicular ordering.

Another conclusion from the present study is that the NIL groove periodicity cannot be  much larger than $\ell_0$;
otherwise the BCP perpendicular order is lost [see Fig.~\ref{fig4}(b) and (c)], where the perpendicular order
is lost for about $d\simeq 20\ell_0$.


Our SCF calculations are in good agreement with the presented
experiments, and provide some insight into the conditions needed to obtain perpendicular lamellar phases
with minimal amount of in-plane defects.
We hope that in the future more detailed three-dimensional calculations as well as careful investigations of film rheology
will shed more light on the fundamental behavior as well as
applications of BCP films in presence of nano-patterned surfaces.

\section*{Acknowledgements}

The help of S. Niedermayer in sample preparation is gratefully acknowledged.
We would like to thank the Triangle de
la Physique, France (POMICO project No.
2008-027T) for supporting the visits of DA and XM
to Saclay. This work
was supported in part by the U.S.-Israel Binational Science Foundation under Grant No.
2006/055, the Israel Science Foundation under Grant No. 231/08, the Center for Nanoscience and Nanotechnology at Tel Aviv
University and the CEA (France)
under programs ``Chimtronique" and ``Nanosciences".



\newpage
\centerline{for Table of Contents use only}
\vskip 2 truecm
\centerline{\bf Organization of Block Copolymers
using NanoImprint Lithography:}
\centerline{\bf Comparison of Theory and Experiments}
\centerline{\it Xingkun Man, David Andelman, Henri Orland, Pascal Th\'ebault,}
\centerline{\it Pang-Hung Liu,Patrick Guenoun, Jean Daillant, Stefan Landis}

\vskip 2truecm

\begin{figure}[h]
\begin{center}
\includegraphics[bb=0 0 403 190, scale=0.7,draft=false]{graphic.eps}
\end{center}
\end{figure}

\end{document}